\documentstyle[12pt,epsfig]{article}

\newlength{\dinwidth}
\newlength{\dinmargin}
\def\lapproxeq{\lower .7ex\hbox{$\;\stackrel{\textstyle
<}{\sim}\;$}}
\def\gapproxeq{\lower .7ex\hbox{$\;\stackrel{\textstyle
>}{\sim}\;$}}

\def\be{\begin{equation}}
\def\ee{\end{equation}}
\def\bea{\begin{eqnarray}}
\def\eea{\end{eqnarray}}

\def\funp{{I\!\!P}}

\def\lapproxeq{\lower .7ex\hbox{$\;\stackrel{\textstyle
<}{\sim}\;$}}
\def\gapproxeq{\lower .7ex\hbox{$\;\stackrel{\textstyle
>}{\sim}\;$}}
\def\be{\begin{equation}}
\def\ee{\end{equation}}
\def\bea{\begin{eqnarray}}
\def\eea{\end{eqnarray}}
\def\funp{{I\!\!P}}
\def\ra{\longrightarrow}
\def\dr{\raisebox{2.1ex}{$\scriptsize\lfloor$}\!\raisebox{1ex}{$\rightarrow$}}
\def\bb{b\bar{b}}

\def\ra{ \rightarrow }

\begin{document}
\titlepage
\begin{flushright}
IPPP/02/29 \\
DCPT/02/58 \\
12 August 2002 \\
\end{flushright}

\vspace*{2cm}

\begin{center}
{\Large \bf Ways to detect a light Higgs boson at the LHC}

\vspace*{1cm}
A. De Roeck$^{a}$, V.A. Khoze$^{b,c}$, A.D. Martin$^b$, R. Orava$^d$ and M.G. Ryskin$^{b,c}$ \\

\vspace*{0.5cm} $^a$ CERN, Geneva\\
$^b$ Institute for Particle Physics Phenomenology, University of
Durham, DH1 3LE, UK \\
$^c$ Petersburg Nuclear Physics Institute, Gatchina,
St.~Petersburg, 188300, Russia \\
$^d$ Department of Physical Sciences, University of Helsinki, and Helsinki Institute of Physics, Finland\\
\end{center}

\vspace*{1cm}

\begin{abstract}
We summarize the possible processes which may be used to search
for a Higgs boson, of mass in the range 114--130~GeV, at the LHC.
We discuss, in detail, two processes with rapidity gaps: exclusive
Higgs production with tagged outgoing protons and production by
Weak Boson Fusion, in each case taking $H\ra\bb$ as the signal. We
make an extensive study of all possible $\bb$ backgrounds, and
discuss the relevant experimental issues. We emphasize the special
features of these signals, and of their background processes, and
show that they could play an important role in identifying a light
Higgs boson at the LHC.
\end{abstract}

\newpage
\section{Introduction}

The identification of the Higgs boson(s) is one of the main goals
of the Large Hadron Collider (LHC) being constructed at CERN.
According to current theoretical prejudice it is likely that a
Higgs boson will exist in the mass range $114<M_H<135$~GeV. In the
Standard Model description of electroweak data, the virtual
effects favour a Higgs with mass at, or just above, the LEP bound
of 114~GeV. Moreover, in the Minimal Supersymmetric Model a scalar
Standard-Model-like Higgs boson with mass below 135~GeV should
exist\footnote{A discussion of, and references for, the current
status of allowed masses and other properties of Higgs bosons can
be found, for example, in Ref.~\cite{HIGGS}.}. However, the
experimental detection of such a `light' Higgs boson at the LHC
will be challenging. There is no obvious perfect detection
process. Rather there is a range of complementary possibilities,
as illustrated in Table~1. The Table shows the number of
identified Higgs events and the number of background events, for a
Standard Model Higgs boson of mass $M_H=120$~GeV for the
integrated luminosity of 30~fb$^{-1}$ planned for the first two or
three years of LHC running, for each of the various proposed
detection channels. A glance at the Table shows that, {\em either}
large signals are accompanied by a huge background, {\em or} the
processes have comparable signal and background rates for which
the number of Higgs events is rather small.
\begin{table}[h]
\begin{center}
\begin{tabular}{l|c|c|c|c}\hline
& \multicolumn{2}{|c|}{number of events}& & significance \\
Higgs signal & signal & background & $S/B$ & $S/\sqrt{S+B}$ \\
\hline


\raisebox{-3ex}{\qquad\qquad\qquad\qquad\qquad\qquad\qquad CMS}  &
\raisebox{-3ex}{313} & \raisebox{-3ex}{5007} &
\raisebox{-3ex}[0pt][5ex]{0.06 ${\displaystyle \left(\frac{1\:{\rm
GeV}}{\Delta M_{\gamma\gamma}}
\right)}$} & \raisebox{-3ex}{$4.3\sigma$} \\
\raisebox{1ex}{a) $H\ra\gamma\gamma$}\qquad\qquad\qquad\qquad
\raisebox{-3ex}{ATLAS} & \raisebox{-3ex}{385} &
\raisebox{-3ex}{11820} & \raisebox{-3ex}[0pt][7ex]{0.03
${\displaystyle \left(\frac{2\:{\rm GeV}}{\Delta M_{\gamma\gamma}}
\right)}$} & \raisebox{-3ex}{$3.5\sigma$} \\
\hline

\raisebox{-3ex}[0ex][3.3ex]{b) $t\bar{t}H $}  &
\raisebox{-3ex}{26} & \raisebox{-3ex}{31} &
\raisebox{-3ex}[0pt][2ex]{0.8 ${\displaystyle \left(\frac{10\:{\rm
GeV}}{\Delta M_{\bb}} \right)}$} & \raisebox{-3ex}{$3\sigma$} \\   $\qquad\, \dr$ \raisebox{1ex}[0ex][-2ex]{$\bb$}&&&& \\
\hline

\raisebox{-3ex}[0ex][3.3ex]{c) $gg^{PP}\ra p+H+p$} &
\raisebox{-3ex}{11} & \raisebox{-3ex}{4} &
\raisebox{-3ex}[0pt][2ex]{3 ${\displaystyle\left(\frac{1\:{\rm
GeV}}{\Delta M_{\rm
missing}}\right)}$} & \raisebox{-3ex}{$3\sigma$} \\
$\qquad\qquad\qquad\ \ \,\dr$ \raisebox{1ex}[0ex][-2ex]{$\bb$}&&&&\\
\hline

\raisebox{-3ex}[0ex][3.3ex]{d) $gg^{PP}\ra X+H+Y$} &
\raisebox{-3ex}[0ex][0ex]{190} & \raisebox{-3ex}[0ex][0ex]{21,000}
& \raisebox{-3ex}[0ex][2ex]{0.009
${\displaystyle\left(\frac{10\:{\rm GeV}}{\Delta M_{\bb}}\right)}$} & \raisebox{-3ex}[0ex][0ex]{$1.3\sigma$} \\
$\qquad\qquad\qquad\quad\;\, \dr$ \raisebox{1ex}[0ex][-2ex]{$\bb$}&&&&\\
\hline

\raisebox{0ex}[3ex][1.9ex]{e) Weak Boson Fusion (WBF) } &  &  &  & \\
$qWWq\ra jHj\ra j\gamma\gamma j$ & 17 & 9 & CMS & $3.3\sigma$ \\
& 18 & 17 & ATLAS & $3\sigma$\\
\qquad\qquad\qquad\ \ $\:\ra j\tau\tau j$& 25 &
8 && $4.4\sigma$ \\
\qquad\qquad\qquad\ \ $\:$\raisebox{0ex}[0ex][3ex]{$\ra
jW(l\nu)W^\ast(l\nu)j$}& 49 & 31 && $5.4\sigma$ \\ \hline

\raisebox{0ex}[3ex][1.9ex]{f) WBF with rapidity gaps } & \multicolumn{2}{|c|}{jet $E_T$ cuts:} &  & \\
{$qWWq\ra j+H({\rm high}\,q_t)+j$} & {250} & {1800} &
\raisebox{0ex}[2ex][1.5ex]{0.14\ {$\displaystyle
\left(\frac{10\:{\rm GeV}}{\Delta
M_{\bb}} \right)$}}  & {$5.5\sigma$} \\
$\qquad\qquad\qquad\ \; $\raisebox{-0.5ex}[0pt][2ex]{$\dr$}
\raisebox{0.5ex}[0pt][2ex]{$\bb$}& \multicolumn{2}{|c|}{Higgs
$q_t$ cut:}&  & \\
& 400 & 3700 & 0.11\ \raisebox{0ex}[0ex][4ex]{$\displaystyle
\left(\frac{10\:{\rm GeV}}{\Delta M_{\bb}} \right)$} & $6.2\sigma$
\\ \hline

\raisebox{0ex}[3ex][1ex]{g) $gg\ra ZZ^{\ast}\ra 4l$} & {6} & {4} & {CMS} & {$1.9\sigma$} \\
& \raisebox{0ex}[0ex][2ex]{3} & {1.5} & {ATLAS} & {$1.4\sigma$} \\
\hline

\raisebox{0ex}[3ex][2ex]{h) $gg\ra WW^{\ast}\ra l\nu l\bar{\nu}$}& {44} & {272} & {CMS} & {$2.5\sigma$}\\
\hline

\raisebox{0ex}[3ex][2ex]{i) $WH\ra l\nu \bb$ } & 161 & 7095 & 0.02 &$1.9\sigma$ \\
\hline
\end{tabular}
\end{center}
\begin{caption}
{\footnotesize The number of signal and background events for
various methods of Higgs detection at the LHC. The significance of
the signal is also given. The mass of the Higgs boson is taken to
be 120~GeV and the integrated luminosity is taken to be 30
fb$^{-1}$. The notation $gg^{PP}$ is to indicate that the gluons
originate within overall colour-singlet (hard Pomeron) $t$-channel
exchanges; see, for example, Fig.~1. The entries for the various
processes are taken, or scaled from the results for 100 fb$^{-1}$
luminosity, from references (a) \cite{Z}, (b) \cite{TT}, (d)
\cite{INC}, (e) \cite{Z,WBF}, (g,h) \cite{Z} and (i) \cite{WH},
where the $K$ factors have been omitted. Processes (c) and (f) are
discussed in detail in Sections~2 and 3, respectively, of this
paper. A detailed study of the NLO contributions to the
irreducible background to the $H\ra\gamma\gamma$ signal shows the
$K$ factor is 0.65~that of the signal \cite{BDS}. Taking these $K$
factors into account the authors find that the significance of the
$H\ra\gamma\gamma$ signal may increase to $7\sigma$.}
\end{caption}
\end{table}

Besides containing the conventional processes for the detection of
a light Higgs boson, the Table also lists processes (c,d,f) with a
rapidity gap on either side of the boson, which provide a clean
environment for its production. These processes are often
overlooked, but they have special advantages. Here we shall study
two of these processes in detail.

First, in Section~2, we discuss the exclusive process $pp\ra p + H
+ p$, where the + sign indicates the presence of a rapidity gap.
We show that it is possible to tag the outgoing protons such that
the Higgs may be identified, and its mass measured to an accuracy
of about 1~GeV, using the `missing mass' method. That is using
tagged protons we have $M_H=M_{\rm missing}$ with $\Delta M_{\rm
missing}\sim 1\:{\rm GeV}$. Importantly, the process allows an
independent measurement of the Higgs mass via the $H\ra\bb$ decay,
$M_H=M_{\bb}$, although now the resolution is much poorer with
$\Delta M_{\bb}\sim\:10{\rm GeV}$. The existence of matching
peaks, centred about $M_{\rm missing} = M_{\bb}$, is a unique
feature of the exclusive diffractive Higgs signal\footnote{This
may be contrasted with the search for a Higgs peak sitting on a
huge background in the $M_{\gamma\gamma}$ spectrum, see
process~(a) of Table~1.}. Besides its obvious value in identifying
the Higgs and in sharpening the determination of its mass, we will
see that the mass equality also plays a key role in reducing
background contributions. Another advantage of the $H\ra \bb$
signal is that, at leading order, the $gg\ra \bb$ background
process is suppressed by a $J_z=0$ selection rule, see Section~2.3
\cite{Liverpool, KMRmm}.

Of course, we have to pay a price for the survival of the rapidity
gaps, so the event rate is low. Nevertheless the process has the
advantage that the signal exceeds the background. The absolute
value of the $pp\ra p+H+p$ cross section has been calculated in
Refs.~\cite{INC,KMR}. The derivation is outlined in Section~2.1.
Moreover the predicted value can be checked experimentally by
measuring the rate of the double-diffractive production of dijets
(of comparable mass). The uncertainties of the calculation of the
$pp\ra p + H + p$ cross section are discussed in Section~2.2.

Equally important to the calculation of the exclusive
double-diffractive $p+(H\ra \bb) + p$ signal, is the estimation of
the $\bb$ background. The background processes are found to be
suppressed by their spin and colour structure, and are interesting
in their own right. The various sources of the background are
discussed in Sections~2.3 and 2.4. First, in Section~2.3, we
summarize the sources, together with their sizes relative to the
Higgs signal, and then, in Section~2.4, we give the detailed
justification of the results. Section~2.5 is devoted to a study of
production by soft Pomeron-Pomeron collisions. We find that it
gives a negligible contribution both to the exclusive $H\ra\bb$
signal and to the background. In Section~2.6, we turn to a
discussion of the experimental issues connected with the exclusive
diffractive signal. We study the experimental efficiencies, the
choice of cuts, the accurate determination of the `missing mass'
via the measurement of the forward protons and, finally, the
problems connected with the `pile-up' of multiple interactions in
each beam crossing.

The second process that we study is the central production of a
Higgs boson via $WW$ fusion
$$pp\ra qWWq\ra {\rm jet} + H + {\rm jet},$$
see Section~3. We may suppress the background to this process by
exploiting the fact that the cross section is rather flat as a
function of the transverse momentum, $q_t$, of the Higgs boson, on
account of the large $W$ boson mass. Moreover, since this process
is mediated by $t$ channel $W$ exchange, which is a point-like
colourless object, there is no corresponding bremsstrahlung in the
central region \cite{TDK,SDK,Bj,FS} and hence no Sudakov
suppression of the rapidity gaps.

\section{Exclusive diffractive $H\ra \bb$ production}

In this section we make a detailed study of the third process of
Table~1, namely \cite{KMR,KMRmm}
$$pp\ra p+H+p$$
where the Higgs decays via the $\bb$ mode. We assume that the
outgoing protons and the $b$ and $\bar{b}$ jets can be identified
and measured, and that there are no other particles in the final
state. We quantify both the signal and the background for this
exclusive process and, moreover, discuss the relevant experimental
issues.

\subsection{Calculation of the $p+H+p$ cross section.}

The basic mechanism for the process is shown in Fig.~1.
\begin{figure}[!h]
\begin{center}
\epsfig{figure=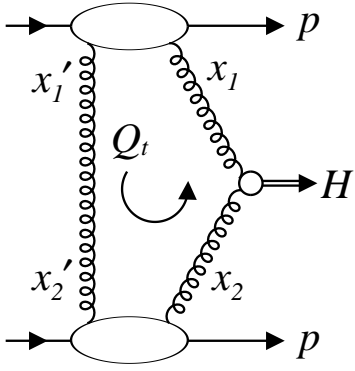,height=2.2in}
 \caption{Schematic diagram of the exclusive double-diffractive production of a Higgs boson, that is the process $pp\ra p+ H + p$, in which the + signs indicate the presence of rapidity gaps.}
 \label{Fig.1}
\end{center}
\end{figure}
It turns out that the typical values of the transverse momentum
$Q_t$ of the gluon, which screens the colour, are much smaller
than $M_H$, but are yet sufficiently large for perturbative QCD to
be applicable. The amplitude is \cite{KMR, KMRmm}
\begin{equation}
{\cal M} = A\pi^3\int\frac{d^2Q_t}{Q_t^4}f_g\left(x_1, x_1^\prime,
Q_t^2, \frac{M_H^2}{4}\right)f_g\left(x_2, x_2^\prime, Q_t^2,
\frac{M_H^2}{4}\right) \label{eq:amplitudeM}
\end{equation}
where the $f_g$'s are the skewed unintegrated gluon densities of
the proton, and where the $gg\ra H$ vertex factor is
\begin{equation}
A^2 =
K\frac{\sqrt{2}G_F}{9\pi^2}\alpha_S^2(M_H^2)\label{eq:Asquared}
\end{equation}
with the NLO $K$ factor $K\simeq 1.5$. The longitudinal momentum
fractions carried by the gluons satisfy
\begin{equation}
\left(x^\prime\sim\frac{Q_t}{\sqrt{s}}\right) \ll
\left(x\sim\frac{M_H}{\sqrt{s}}\right) \ll 1
\end{equation}
where, for the LHC, $\sqrt{s}=14$~TeV. In this domain the skewed
unintegrated densities are given in terms of the conventional
(integrated) densities $g(x,Q_t^2)$. To single log accuracy, we
have \cite{KMR}
\begin{equation}
f_g\left(x,x^\prime,Q_t^2, \frac{M_H^2}{4}\right) =
R_g\frac{\partial}{\partial\ln Q_t^2}\left(\sqrt{T\left(Q_t,
\frac{M_H}{2}\right)}xg\left(x,Q_t^2\right)\right)
\label{eq:fsubg}
\end{equation}
where $T(Q_t, \mu)$ is the survival probability that the gluon
remains untouched in the evolution up to the hard scale $\mu =
M_H/2$. This Sudakov factor $T$ is the result of resumming the
virtual contributions in the DGLAP evolution. It is given by
\begin{equation}
T(Q_t,\mu) =
\exp\left(-\int_{Q_t^2}^{\mu^2}\frac{\alpha_S(k_t^2)}{2\pi}\frac{dk_t^2}{k_t^2}
\int_0^{\mu/(\mu+k_t)}\left[zP_{gg}(z) + \sum_q
P_{qg}(z)\right]dz\right). \label{eq:T}
\end{equation}

The square root of $T$ arises in (\ref{eq:fsubg}) because the
survival probability is only relevant to the hard gluon. Note that
the gluon with $x^\prime \simeq 0$ is almost `at rest' and so
there is no possibility of QCD radiation \cite{MR}. The
multiplicative factor $R_g$ is the ratio of the skewed
$x^\prime\ll x$ integrated distribution to the conventional
diagonal density. For $x\ll 1$ the factor is completely determined
\cite{SGMR}. We find $R_g \simeq 1.2$ at the energy of the LHC.

The radiation associated with the $gg\ra H$ hard subprocess is not
the only means to populate and to destroy the rapidity gaps. There
is also the possibility of soft rescattering in which particles
from the underlying event populate the gaps. The probability,
$S^2$, that the gaps survive the soft $pp$ rescattering was
calculated using a two-channel eikonal model, which incorporates
high mass diffraction \cite{KMRsoft}. The parameters of the model
were obtained from a global analysis of all available soft $pp$
(and $p\bar{p}$) scattering data. In this way, we find $S^2=0.02$
for the process $p+H+p$ at the LHC. Including this factor, the
cross section is predicted to be \cite{INC}
\begin{equation}
\sigma(pp\ra p+H+p)\simeq 3\:{\rm fb} \label{eq:sigma}
\end{equation}
for the production of a Standard Model Higgs of mass 120~GeV at
the LHC.

In Section~2.2 we estimate a factor of two uncertainty in the
cross section prediction given in (\ref{eq:sigma}). On the other
hand it is frequently quoted that the predictions of the cross
section for diffractive Higgs production cover many orders of
magnitude, and for this reason the many authors choose not to
consider this Higgs signal. This is unfortunate. Sometimes the
rates of different diffractive mechanisms are compared. Sometimes
models are used to predict the exclusive signal which are not
valid. Indeed, care must be taken when comparing the theoretical
predictions for the exclusive $pp\ra p+H+p$ process with the
results of Monte Carlo simulations. For example, when implementing
the Soft Colour Interaction (SCI) prescription \cite{SCI} in
PYTHIA \cite{PYTHIA}, it was found \cite{SCITev} that hard
production in single diffractive processes observed at the
Tevatron could be described reasonably well, but that the
generator hardly ever produces any `double-Pomeron-exchange'
events. Finally, an extremely low limit was claimed \cite{SCIH}
for the exclusive $pp\ra p+H+p$ cross section.

The fact that such a generator yields an extremely low probability
for exclusive processes is not surprising. The generator was
created to simulate {\em inelastic} processes. It operates by
starting from the hard subprocess and generates the parton showers
by backward evolution. The generator never accounts for the
important coherence between different parton showers, nor for the
colourless nature of the initial particles. The incoming protons
are just considered as a system of coloured partons. As a
consequence, the probability not to emit additional secondary jets
(and so to reproduce an exclusive process) turns out to be
negligibly small. In particular, such a generator is unable to
reproduce the elastic cross section. Such generators create many
secondary minijets at the parton shower stage and the probability
to screen all these minijets by colour interchange is extremely
low. Such generators were not constructed to reproduce exclusive
processes, where the colour coherence effects and colourless
nature of the incoming hadrons are important.

\subsection{Uncertainty in the cross section prediction}

Note that, in principle, (\ref{eq:sigma}) is an absolute
prediction for $\sigma(p+H+p)$. Of course, the various inputs are
subject to uncertainty. Let us discuss these in turn. First we
have the large suppression from the probability $S^2 = 0.02$ that
the rapidity gaps survive soft $pp$ rescattering. From the
analysis \cite{KMRsoft} of all soft $pp$ data we estimate the
accuracy of the prediction for $S^2$ is $\pm50\%$. One check of
the eikonal model calculations of $S^2$ is the estimate of the
diffractive dijet production rate measured by the CDF
collaboration \cite{CDFjj} at the Tevatron. The rate, when
calculated using factorization and the diffractive structure
functions obtained from HERA data, lies about a factor of 10 above
the CDF data. However, when rescattering corrections are included,
and the survival probabilities computed, remarkably good agreement
with the CDF measurements is obtained \cite{KKMR}. There is,
perhaps, a small tendency that an even stronger suppression is
required, so the true survival probabilities $S^2$ may be a bit
smaller than our predictions.

Second, although, on account of the Sudakov form factor, the $Q_t$
integral in (\ref{eq:amplitudeM}) is infrared safe, the cross
section may have some contribution from the non-perturbative
region. Again, we expect an accuracy of $\pm50\%$, but in this
case the + sign looks more realistic. The uncertainties in the
gluon densities $f_g$ in the integrand are estimated at $\pm5\%$,
leading to a $\pm22\%$ error on the cross section. This estimate
takes into account the accuracy of the value of $R_g$ and the
uncertainty on the integrated gluon density $xg$, at $x\sim0.01$
relevant to the LHC. Finally, we have NLO contributions to the
Sudakov $T$ factor ($\pm20\%$) and NNLO corrections to the $gg\ra
H$ vertex factor ($\pm20\%$). Adding these errors in quadrature
gives a {\em factor two} uncertainty in (\ref{eq:sigma}).

We stress that the predicted value of the cross section can be
checked experimentally. All the ingredients, except for the NLO
correction to the $gg\ra H$ vertex, are the same for our signal as
for exclusive double-diffractive dijet production, $pp\ra p+ {\rm
dijet} + p$, where the dijet system is chosen in the same
kinematic domain as the Higgs boson, that is $M(jj)\sim120\:{\rm
GeV}$ \cite{INC,KMR}. Therefore by observing the larger dijet
production rate, we can confirm, or correct, the estimate of the
exclusive Higgs signal.

Of course, so far our discussion has been within the confines of
the Standard Model. We should not overlook the possibility that
the exclusive Higgs signal may reveal New Physics. For example,
the Higgs cross section will be 4 (or even 9) times larger if
there were to exist a fourth doublet containing one (or two)
quarks heavier than the Higgs.

\subsection{Summary of the backgrounds to the \ $p+(H\ra \bb)+p$ \ signal}

For a light Higgs boson the dominant decay channel is $H\ra\bb$.
However, it is impossible to observe this decay in an inclusive
process at the LHC, since it is overwhelmed by a huge QCD $\bb$
background. The advantage of exclusive double-diffractive Higgs
production with forward going protons is that there exists a
$J_z=0$, parity even, selection rule, which strongly suppresses
the leading order $gg^{PP}\ra\bb$ background subprocess. The $PP$
superscript is to indicate that each gluon comes from a
colour-singlet $t$ channel state. One way to see the $J_z=0$
selection rule is to note that, in the equivalent gluon approach
(or in a planar gauge), the polarisation vectors of the $t$
channel gluons in Fig.~1 are aligned along the transverse
component of the loop momentum $\vec{Q}_t$ \cite{INC,KMRmm}.
Integrating over $\vec{Q}_t$ leads to the amplitude ${\cal M}$ of
the $gg^{PP}$ fusion subprocess being formed from the average over
the two transverse polarisations $\varepsilon_1,\varepsilon_2$ of
the incoming gluons
\begin{equation}
{\cal M} = \frac{1}{2}\sum_{\varepsilon_1, \varepsilon_2}{\cal
M}(\varepsilon_1, \varepsilon_2)\delta_{\varepsilon_1,
\varepsilon_2}. \label{eq:matrixelementsM}
\end{equation}
The delta symbol reflects the $J_z=0$, parity even, character of
the di-gluon $gg^{PP}$ incoming state. For {\em colour-singlet}
$\bb$ production the Born-level contributions of Figs.~2(a) and
2(b) cancel each other, in the massless $m_b\ra0$ limit, due to
summation (\ref{eq:matrixelementsM}). More generally, as a
consequence of helicity conservation, and $P$ and $T$ invariance,
for the real parts of the $J_z = 0$\ $gg^{PP}\ra\bb$ amplitudes
\cite{BORDEN}, the corresponding contributions to the cross
section vanish with decreasing quark mass as $m_b^2/E_T^2$, where
$E_T\sim M_H/2$ is the transverse energy of the $b$ or $\bar{b}$
jet.
\begin{figure}[!h]
\begin{center}
\epsfig{figure=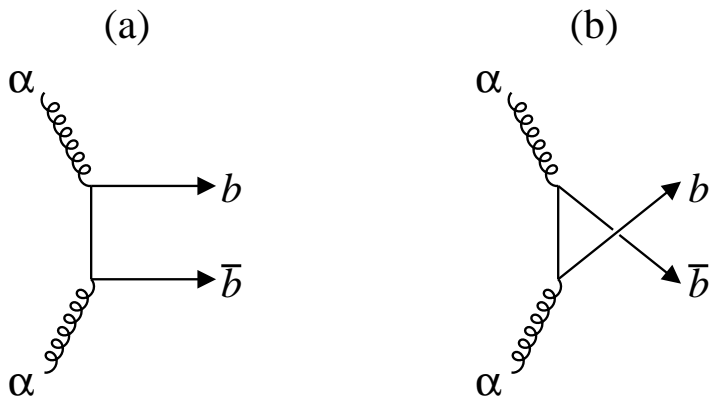,height=2.2in}
 \caption{Colour-singlet $gg\ra\bb$ production, where $\alpha$ denotes the colour of the incoming gluons.}
 \label{Fig.2}
\end{center}
\end{figure}

It is convenient, first, to list the possible sources of the $\bb$
background to the exclusive $p+(H\ra\bb)+p$ Higgs signal and to
state the size of each background in terms of the signal. The
justification for the numerical values is given in the next
subsection,~2.4. We quote the $B/S$ using the anticipated missing
mass resolution, $\Delta M_{\rm missing} = 1\:{\rm GeV}$, expected
from employing taggers for the outgoing protons.

At leading order (LO) there are $gg^{PP}\ra $``$\,\bb\,$''
background contributions, despite the $J_z=0$ selection rule,
which we summarize below.

\begin{itemize}
\item[(i)] The prolific $gg^{PP}\ra gg$ subprocess may mimic
$\bb$ production since we may misidentify the outgoing gluons as
$b$ and $\bar{b}$ jets. Assuming the expected 1\% probability of
misidentification and applying $60^\circ < \theta < 120^\circ$ jet
cut, we estimate $B/S\sim0.06$.
\item[(ii)]There is an admixture of $|J_z|=2$ production, arising
from non-forward going protons \cite{KMRmm}. It gives
$B/S\sim0.08$.
\item[(iii)] For a massive quark there is a contribution to the $J_z=0$ LO cross section of order $m_b^2/E_T^2$, leading to $B/S\sim0.06$.
\end{itemize}

At next-to-leading order(NLO), we have the possibility of
$gg^{PP}\ra \bb g$ background contributions.

\begin{itemize}
\item[(iv)] The extra gluon may go unobserved in the direction of
a forward proton. This background may be effectively eliminated by
requiring the approximate equality $M_{\rm missing} = M_{\bb}$.
\item[(v)] The extra gluon is collinear with either the $b$ or
$\bar{b}$ jet. We will show that this is suppressed for soft
radiation by the specific spin structure of the
process\footnote{This extension of the $J_z=0$ selection rule
suppression to soft radiation is true for soft radiation at any
angle.}, and leaves a background coming from three jet $\bb g$
production with $B/S\sim0.06$.
\end{itemize}

We also consider NNLO $\bb gg$ background contributions.

\begin{itemize}
\item[(vi)] There is a NNLO $\bb$ contribution, which comes from
the diagram formed by the product of the NLO
amplitudes, which is negligibly small.
\item[(vii)] Another source of this background may be called
central inelastic production \cite{INC}, where the $H\ra\bb$
signal or QCD $\bb$ background is accompanied by central soft
gluon radiation. We will show that it causes a 1--2\% high mass
tail to the `missing mass' Higgs signal, and again gives a
negligible contribution to the background.
\end{itemize}

Finally, we consider the effects from the collisions of two soft
Pomerons. After imposing the missing mass equality, we find that
production by Pomeron-Pomeron fusion gives a negligible
contribution to both the $H\ra\bb$ signal and the $\bb$
background. Pomeron-Pomeron fusion is discussed separately in
Section~2.5.

In summary, taking all these sources of background into account,
we would expect a signal/background ratio of about four. The
reliability of this signal-to-background estimate, $S/B\simeq4$,
with respect to the theoretical uncertainties, is much better than
the factor of two uncertainty in the signal itself. This is
because the gap survival probability, $S^2$, the unintegrated
gluon distributions, $f_g$, the NLO $T$-factor and even the
contribution from the low $Q_t$ non-perturbative domain are
common\footnote{Except for the $|J_z|=2$ admixture which has a
different structure in the $Q_t$ loop integral for the signal and
the background.} to both the signal, $S$, and the background, $B$.
Thus the main theoretical uncertainties cancel in the ratio.
However, there is another source of uncertainty due to the
higher-order virtual corrections to the background processes. We call
these, at present unknown, contributions the background
$K$~factor. Since we have included the $K$ factor for the Higgs
signal in (\ref{eq:Asquared}), we conservatively take the same $K$
factor for the background. The effect is to reduce $S/B$ to about
3.

Of course, the experimental uncertainties depend on the values of
the mass resolutions, $\Delta M_{\rm missing}$ and $\Delta
M_{\bb}$, the probability to misidentify a gluon as a $b$-jet, the
$b$-jet tagging efficiency and on the appropriate choice of the
jet cone size $\Delta R$; see Section~2.6. Here we have taken
values which should be attainable, but clearly, in practice, these
have to be optimized with reference to the detector, and further
gains are possible.

\subsection{Determination of the individual backgrounds}

Here we quantify each of the potential backgrounds listed above.
As a reference point we start with the largest exclusive signal,
that is one in which we have no $b$ jet identification. In other
words, we have the central production of a pair of high $E_T$ jets
together with tagged outgoing protons. For this situation the
ratio of the background to the Higgs signal\footnote{Note that the
NLO $K$ factor is included for the $gg\ra H$ vertex,
(\ref{eq:Asquared}), but in (\ref{eq:BtoS}) we omit
the virtual NLO corrections to the background. For
presentation purposes, the denominator in
(\ref{eq:BtoS}) actually refers to only the $H\ra\bb$ component
and does not include the $H\ra gg$ decays; for $M_H=120$~GeV,
including the $H\ra gg$ mode would enlarge the denominator by
about 10\%.} is
\begin{equation}
\frac{B(gg^{PP}\ra gg)}{S(gg^{PP}\ra H\ra jj)}\simeq
600\left(\frac{\Delta M}{1\:{\rm GeV}}\right),\label{eq:BtoS}
\end{equation}
where we have normalised the ratio to the expected experimental
missing mass resolution, $\Delta M_{\rm missing} = 1.0\:{\rm
GeV}$. For clarity, we omit this factor in brackets from now on in
this subsection. In (\ref{eq:BtoS}), the QCD background has been
suppressed by imposing the polar angle acceptance cut,
$60^\circ<\theta<120^\circ$, in the jet-jet centre-of-mass frame.
The cut removes half of the $\bb$ Higgs signal, but more
importantly removes the $gg^{PP}\ra gg$ infrared singularity.

Using (\ref{eq:BtoS}) as a reference point, we now estimate the
background/signal for the exclusive process for each of the
sources of background listed in Section~2.3. We divide them into
background contributions at LO, NLO and NNLO.

\subsubsection{The LO $\bb$ backgrounds}

\begin{itemize}
\item[(i)] Clearly, for the exclusive double-diffractive Higgs signal to be of value, we must reduce $B/S$ of (\ref{eq:BtoS}). We therefore need to tag both the $b$ and $\bar{b}$ jets. Even so,
there is a chance that the gluons are misidentified as $b$ and
$\bar{b}$ jets. The expected probability of misidentification is
about 1\%. Therefore by observing the $b$ and $\bar{b}$ jets we
reduce the background by $10^4$, and hence
\begin{equation}
\frac{B(gg^{PP}\ra gg\ra {\rm ``}\,\bb\,{\rm ''})}{S(gg^{PP}\ra
H\ra \bb)}\simeq 0.06.\label{eq:BtoS0.06}
\end{equation}
\item[(ii)] Of course, there is a background from QCD $\bb$
production itself. We have emphasized that at LO this vanishes in
the massless quark limit, $m_b\ra0$, due to the $J_z=0$ selection
rule. However, there is an admixture of $|J_z|=2$ caused by the
transverse momenta $\vec{p}_{it}$ of the outgoing protons. In the
exact forward direction, the $J_z=0$ selection rule is simply a
consequence of angular momentum conservation and the absence of
spin-spin correlations between particles separated by a large
rapidity gap\footnote{Here we refer to the correlation between
`two spin-flips'. It is very small, and moreover decreases rapidly
with beam energy, in soft processes. At small distances it is
described by the spin structure function $g_2(x)$, which, in
comparison with the unpolarised structure function, is suppressed
by a power of $x$. That is the correlation is suppressed
exponentially by the size of the rapidity gap.}. However,
violation of the rule can come from orbital angular momenta,
$p_{it}r$. For our process, the distance $r$ is controlled, not by
the size of the proton, but by the effective size of the
$t$-channel $gg$ state. Hence $r\sim 1/Q_t$. Therefore the
admixture of $|J_z|=2$ states is strongly suppressed by the ratio
$4p_{1t}^2p_{2t}^2/Q_t^4$. It was estimated in \cite{KMRmm} that
the mean $|J_z|=2$ admixture is less than 1.5\%. In addition,
$\bb$ production (even for $|J_z|=2$) is suppressed in comparison
to $gg$ production by a factor $27\times4$ due to the colour and
spin $\frac{1}{2}$ character of the quark. The factor represents
the exact ratio of the subprocess cross sections at
$\theta=90^\circ$ (see equations (49) and (52) of \cite{INC}).
Thus we have
\begin{equation}
\frac{B_{|J_z|=2}(gg^{PP}\ra\bb)}{S(gg^{PP}\ra H\ra\bb)}\equiv
\frac{\bar{B}}{S}\times\left(|J_z|=2\ {\rm admix.}\right)\sim
\frac{\bar{B}}{S}\times0.015\sim
\frac{600\times0.015}{27\times4}\sim 0.08,\label{eq:BbartoS}
\end{equation}
where $\bar{B}$ would have been the $gg^{PP}\ra\bb$ background if
the $J_z=0$ selection rule had not existed, that is if we had
averaged over the incoming gluon helicities in the usual way (as
for inclusive production), rather than as in
(\ref{eq:matrixelementsM}).
\item[(iii)] A second way to avoid the $J_z=0$ prohibition of the
LO background process, $gg^{PP}\ra\bb$, is to allow for the $b$
quark mass. At LO the cross section is suppressed by a factor
$m_b^2/E_T^2$ in comparison to $\bb$ production $\bar{B}$ in the
absence of the $J_z=0$ selection. It is even smaller if we account
for the non-Sudakov double-logarithmic suppression \cite{FKM}. For
the acceptance cut $60^\circ<\theta<120^\circ$ the jet transverse
energy $E_T\gapproxeq10m_b$, and thus
\begin{equation}
\frac{B_{m_b}(gg^{PP}\ra\bb)}{S(gg^{PP}\ra
H\ra\bb)}\lapproxeq\frac{600\times10^{-2}}{27\times4}\sim0.06.\label{eq:BmbtoS}
\end{equation}
\end{itemize}

\subsubsection{The NLO $\bb g$ backgrounds}

The NLO subprocess $gg^{PP}\ra\bb g$ also generates a background
for our exclusive double-diffractive Higgs signal. First recall
that the virtual NLO $\alpha_S$ correction has already been
included in the $gg\ra H$ vertex, (\ref{eq:Asquared}). Of course,
the extra gluon jet may be observed experimentally in the central
detector and so these background events can be readily eliminated.
The exceptions are the emission of the gluon, first, in one of the
beam directions and, second, at large angles, either with small
energy $\omega\ll E_T$ or in the $b$ or $\bar{b}$ jet direction.
\begin{itemize}
\item[(iv)] Extra gluon radiation in either of the beam
directions cannot be observed directly. On the other hand we know
it must be energetic. The size of the initial colour-singlet
system is $1/Q_0 < 1\:{\rm GeV}^{-1}$. Therefore the transverse
momentum of the emitted gluon should be greater than 1~GeV. If the
gluon is to go unobserved outside the calorimeter, that is to have
$\eta>5$, then the gluon energy $E=p_t\cosh\eta>75$~GeV. This
considerably violates the required equality $M_{\rm missing} =
M_{\bb}$ for a Higgs signal, and so this background can be
eliminated.
\item[(v)] Now we come to the background associated with large angle gluon
emission. First we note that soft gluon emission from the $b$ and
$\bar{b}$ quark jets themselves is not a problem, since for this
case we retain the cancellation between the graphs of Fig.~2. Next
we note that emission from the virtual $b$ quark line is
suppressed by at least a factor of $\omega/E$ in the amplitude
(and maybe an even higher power\footnote{An explicit calculation
for the $\gamma\gamma(J_z=0)\ra\bb g$ cross section shows a
$(\omega/E)^4$ suppression, see for example \cite{BORDEN}. This is
a consequence of the Low-Burnett-Kroll theorem \cite{LBK}, see
\cite{DKS}.} of $\omega/E$), where $\omega$ and $E$ are,
respectively, the energies of a gluon and outgoing $b$ quark in
the $gg^{PP}\ra\bb$ centre-of-mass frame. Formally, the factor
arises from the large virtuality of the extra $b$ quark
propagator, and reflects the fact that the $1/\omega$ formation
time of the extra soft gluon is much larger than the lifetime of
the virtual $t$ channel $b$ quark.
\begin{figure}[!h]
\begin{center}
\epsfig{figure=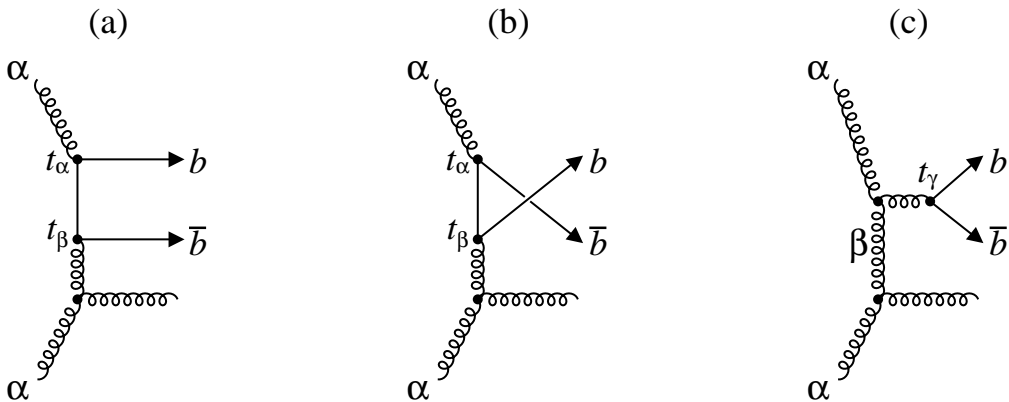,height=2.2in}
 \caption{Colour-singlet $gg\ra\bb g$ production, where $\alpha$, $\beta$ and $\gamma$ are gluon colour labels and where the $t_i$ are the colour matrices for the quark-gluon vertices.}
 \label{Fig.3}
\end{center}
\end{figure}

The potential danger is emission from one of the incoming $t$
channel gluons. At first sight it appears to be a rather large
contribution of order $(\alpha_S N_C/\pi)\ln(M_H/Q_0)$, where the
large log comes from the $d\omega/\omega$ integration embracing
soft gluon emission. On the other hand, to accuracy $\omega/E$, we
deal with pure classical emission which does not change the spin
structure of the amplitude. Therefore we might expect the same
cancellation between the graphs of Fig.~3(a) and 3(b) as was
obtained in Fig.~2. But now the $\bb$ system is in a {\em colour
octet} state and the commutator
\begin{equation}
\left[t_\alpha,t_\beta\right] = if_{\alpha\beta\gamma}t_\gamma
\label{eq:commutator}
\end{equation}
spoils the cancellation. Here $t_i$ are the colour matrices of the
quark-gluon vertices. All is not lost, however, since now we have
a third diagram, Fig.~3(c), which has precisely the colour and
spin structure to restore the cancellation. Thus soft gluon
emissions from the initial colour-singlet $gg^{PP}$ state
factorize (see, for example, \cite{FACT}) and, due to the
overriding $J_z=0$ selection rule, QCD $\bb$ production is still
suppressed. In this way the $\ln(M_H/Q_0)$ contribution is
neutralised and we are left with $O(\alpha_S)$ large angle hard
gluon emission. Such three jet $\bb g$ production can be observed
and excluded experimentally, except for hard gluon radiation along
the $b$ and $\bar{b}$ jet directions. If we denote the cone angle
needed to separate the $g$ jet from the $b$ (or $\bar{b}$) jet by
$\Delta R$, then the expected background from unresolved three jet
events is of about\footnote{We thank Andrei Shuvaev for
calculating this higher-order contribution (see also
Ref.~\cite{DIXON}, where the formalism to obtain
the helicity amplitudes for the $gg\ra\bb
g$ subprocess is given).}
\begin{equation} \left(\alpha_SC_F/2\pi\right)(\Delta
R)^2\bar{B}(gg^{PP}\ra\bb) \ \sim\  0.01\bar{B}(gg^{PP}\ra\bb)
\label{eq:expectedbackground}
\end{equation}
for $\Delta R\sim0.5$, where $C_F = (N_C^2-1)/2N_C = 4/3$. In
(\ref{eq:expectedbackground}) we sum the contributions from gluon
emission in the directions of both $b$ and $\bar{b}$ jets. Recall
that $\bar{B}$ is the $gg^{PP}\ra\bb$ rate in the absence of the
$J_z=0$ selection rule. Noting (\ref{eq:BtoS}) and
(\ref{eq:BbartoS}), we thus see that we have a
background-to-signal ratio of
\begin{equation}
\frac{B(gg^{PP}\ra\bb g)}{S(gg^{PP}\ra H\ra\bb)} \sim
\frac{0.01\times600}{27\times4} \sim 0.06. \label{eq:BtoS0.25}
\end{equation}
This source of background may be further suppressed by choosing
smaller size $b$ and $\bar{b}$ jet cones, $\Delta R$. The price is
the presence of an extra Sudakov form factor which accounts for
the absence of bremsstrahlung outside the more confined cone, that
is radiation which would normally be allowed in the jet
hadronisation. As we need to exclude only hard gluon radiation,
this is a single, not double, logarithmic form factor. It is
present in both the signal and the background, and so does not
change the $S/B$ ratio. However, it reduces the number of selected
events. The form factor is estimated to be $\sim\Delta R^{0.8}$.
On the other hand it is possible that if we take a smaller $\Delta
R$ then we will improve the efficiency of the $b$ and $\bar{b}$
identification. In this way, the Sudakov suppression arising from
a smaller $\Delta R$ could be partly compensated. Of course, in
practice, we should choose the jet cone size $\Delta R$ to
optimize the significance of the signal.
\end{itemize}

\subsubsection{The NNLO $\bb gg$ background}

\begin{figure}[!h]
\begin{center}
\epsfig{figure=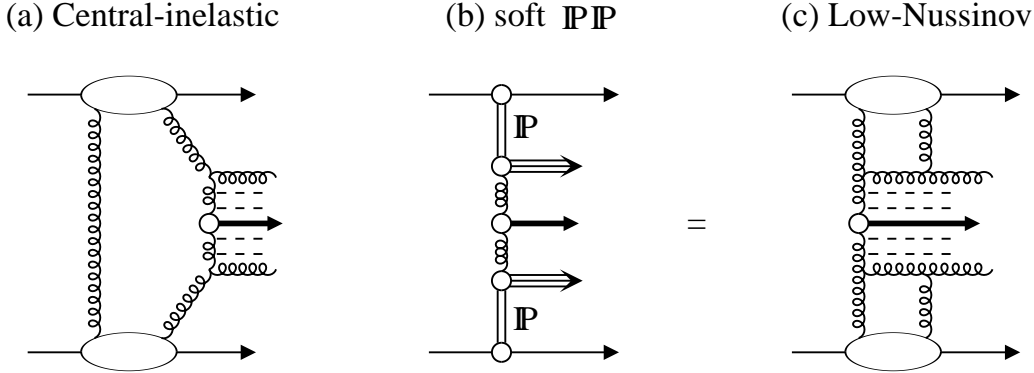,height=2.2in}
 \caption{Double-diffractive production of a Higgs boson (shown by the {\em bold} central arrow) accompanied by gluon emission in diagram~(a) and by Pomeron remnants in diagram~(b). Diagram~(c) shows the Pomeron-Pomeron production process from a QCD viewpoint, in which each Pomeron is represented by two-gluon exchange.}
 \label{Fig.4}
\end{center}
\end{figure}

\begin{itemize}
\item[(vi)]There is a contribution to the $\bb$ QCD background cross
section at NNLO, which comes from the `square' of NLO amplitudes,
in which the $J_z=0$ amplitude for the $gg^{PP}\ra\bb$ subprocess
does not vanish, even in the massless quark limit. Note that, for
the imaginary part of the one-loop box diagram, the arguments
based on $T$ invariance are redundant~\cite{BORDEN}. However,
numerically, this $\bb$ background contribution appears to be very
small (see~\cite{JT} for an explicit calculation in the case of
$\gamma\gamma(J_z=0)\ra q\bar{q}$, and also \cite{FKM}). This is
because it is suppressed by the two-loop factor
$(N_C\alpha_S/4\pi)^2$, where $\alpha_S$ is taken at a large scale
of the order of $M_H/2$. Therefore the corresponding ratio $B/S$
does not exceed 0.01, and can be safely neglected.
\item[(vii)] Finally, there is a background to $pp\ra p+(H\ra\bb)+p$ due
to central inelastic production, where the $\bb$ pair (or Higgs
boson) is accompanied by soft QCD radiation in the central region,
see Fig.~4(a). Hard radiation is excluded by requiring the mass
equality $M_{\rm missing} = M_{\bb}$, as noted above. Moreover,
recall that soft radiation from the final $b$ or $\bar{b}$ quark
lines is already included in the jet finding algorithm. So we are
left with central soft radiation from the $t$ channel lines. Due
to factorization of soft gluon radiation from the colour-singlet
$gg^{PP}$ state (see, for example, \cite{FACT}), this emission
does not alter the signal-to-background ratio. However, it could
blur out the sharp missing-mass peak of the Higgs signal.
Fortunately, the phase space for such radiation is strongly
limited by requiring the $M_{\rm missing} = M_{\bb}$ mass balance.
Consider the emission of two extra soft gluons, that is the  NNLO
subprocess $gg^{PP}\ra Hgg$. (One extra gluon cannot be produced
together with the Higgs boson from the colour-singlet $gg^{PP}$
state, and the $\bb g$ QCD background has already been considered
in (iv) above.) Using factorization, the probability of the
emission of the extra two gluons is
\begin{equation}
\left(\frac{1}{N_C^2-1}\right)\frac{1}{2!}\left[2\left(\frac{\alpha_SN_C}{\pi}\right)\int\frac{d\omega}{\omega}\frac{dp_t^2}{p_t^2}\right]^2,
\label{eq:probability}
\end{equation}
where the first factor reflects colour-singlet production, $1/2!$
accounts for the identity of the two gluons, the factor $2^2$
allows for emission from both incoming gluons and the familiar
double-log accounts for emission with gluon energies $\omega > p_t
> Q_0$, as discussed before. The energy $\omega$ may be bounded by
experimentally requiring $M_{\rm missing} = M_{\bb}$. If we assume
an experimental resolution $\Delta M_{\bb} = 10\:{\rm GeV}$ and
take $\alpha_S = 0.25$ at the low $Q_0$ scale, then the
probability (\ref{eq:probability}) is about 1--2\%. This $Hgg$
background does not affect the significance of the exclusive Higgs
signal, but produces a small tail on the high side of the missing
mass peak. Due to the factorization of soft gluon emission, which
we discussed in (v), the direct QCD $\bb gg$ production is
additionally suppressed by the $J_z=0$ selection rule, and is
hence negligible.
\end{itemize}

\subsection{Production by soft Pomeron-Pomeron collisions}

In addition to the central inelastic production, Fig.~4(a),
studied above, there exists another class of diagrams,
Figs.~4(b~or~c), in which the colour flow is screened in a
different way. These diagrams describe production by soft
Pomeron-Pomeron inelastic collisions\footnote{From a QCD viewpoint
the soft Pomeron-Pomeron interaction, Fig.~4(b), should be
regarded as Fig.~4(c) where the soft Pomerons are replaced by
(Low-Nussinov) two-gluon exchange. We note that diagram~4(c)
contains an extra factor of $\alpha_S$, as compared to
diagram~4(a). Of course the coupling is taken at a low scale, but
nevertheless we should not be surprised when we find the
contribution of 4(c) is less than that of 4(a).}. Within this
approach, one may use the factorizable, a la Ingelman-Schlein \cite{IS},
model in which the Higgs (or $\bb$ pair) is created by a $gg\ra H$
(or $gg\ra\bb$) subprocess that is driven by the `gluon structure
functions' of the Pomerons themselves \cite{PP1,PP2,BDPR}.

Here we check the size of the background to the exclusive Higgs
signal which comes from soft Pomeron-Pomeron collisions. We use
the Donnachie-Landshoff parameterization \cite{DL} to calculate
the Pomeron flux and take the gluon structure function of the
Pomeron to be $zg^P(z)\leq 0.7$, which is consistent with the H1
analysis \cite{H1}. Besides the relatively low values of the
Pomeron flux and of $g^P$, the main suppression comes from the
requirement that the Pomeron-Pomeron mass ($M_{PP} = M_{\rm
missing}$) measured by the tagged protons, should lie within the
$M_{\bb} + \Delta M_{\bb}$ mass interval. This mass balance
requires that the gluon in the Pomeron has momentum fraction $z$
close to 1, where the structure function $g^P(z)$ becomes small.
To allow for uncertainties in the H1 analysis we conservatively
take $g^P(z) = 0.7$, even in this large $z$ domain. Then the cross
section for the Pomeron-Pomeron fusion subprocess is, see
Ref.~\cite{INC}
\begin{eqnarray}
\sigma(\funp\funp\ra\bb) & = & \int dz_1 g^P(z_1)\int dz_2
g^P(z_2)\hat\sigma(gg\ra\bb)\nonumber\\
& \lapproxeq & (0.7)^2\:2\left(\frac{\Delta
M_{\bb}}{M_{PP}}\right)^2\:\hat\sigma(gg\ra\bb).
\end{eqnarray}
We mentioned that the phenomenological flux of the soft Pomeron is
relatively small. Indeed, for $M_{PP} = 120\:{\rm GeV}$ the
effective Pomeron-Pomeron luminosity is a factor two smaller than
the corresponding luminosity for the exclusive process, which was
calculated in terms of the unintegrated gluon distributions
(compare the curves denoted by `soft $\funp\funp$' and `excl' in
Fig.~2(c) of Ref. \cite{INC}). So, finally, the suppression factor
in going from the exclusive process to Fig.~4(b) is
\begin{equation}
\frac{1}{2}\left(\frac{\Delta M_{\bb}}{M_{PP}}\right)^2 \lapproxeq
4\times 10^{-3},\label{eq:suppressionfactor}
\end{equation}
if we take $\Delta M_{\bb} = 10$~GeV. Even though the $J_z=0$
selection rule is absent for Pomeron-Pomeron production, it gives
a background which is much less than the background caused by the
$|J_z|=2$ admixture of (ii).

If we consider Pomeron-Pomeron production of the $H\ra\bb$ signal,
then there is an additional suppression coming from a factor
$1/\left(2(N_C^2-1)\right)$, since the gluons producing the Higgs
must have the same helicity and colour. In practice such a soft
Pomeron-Pomeron source of the signal appears to be very small,
even for $M_{PP}\gg M_H$.

Our estimates of the inclusive double-diffractive background $\bb$
production, when translated to Tevatron energies, agree reasonably
well with the Monte Carlo simulation of Ref.~\cite{PP1}. Of
course, in Ref.~\cite{PP1} only the contribution of Fig.~4(b) was
considered, but the result was normalized to the CDF data
\cite{CDFCox} and in this way the major contribution coming from
Fig.~4(a) type of diagrams was accounted for. On the other hand,
we do not reproduce the results of Ref.~\cite{PP2}.

\subsection{Experimental issues concerning $pp\ra p+H+p$}

The most prominent characteristic feature of diffractive Higgs
production is the formation of  rapidity gaps between the Higgs
decay products and the scattered protons. The gaps will be however
of limited use at the LHC collider, when the machine is operated
at medium ($10^{33}\:{\rm cm}^{-2}{\rm s}^{-1}$) and high
($10^{34}\:{\rm cm}^{-2}{\rm s}^{-1}$) luminosity, due to pile-up,
as discussed in Section~3.2. To select these events in the
experiment it is important to tag the scattered protons. This has
furthermore the crucial advantage that the mass of the Higgs
particle can be precisely reconstructed from the missing mass to
the protons, as detailed below. Due to the relatively low mass of
the central (Higgs) system, the scattered protons have small $\xi$
values, in the range of $10^{-3}$--$10^{-2}$, where $\xi$ is the
momentum fraction lost by the proton in the interaction. A
classical technique to detect scattered protons at small $t$ and
with small relative momentum loss, is by using so-called Roman Pot
detectors. Recently a new type of detectors, called
microstations~\cite{nomokov}, has been proposed for this purpose.
Studies of the  LHC beam optics~\cite{orava} reveal that, in order
to access these small $\xi$ values, the Roman Pot detectors or
microstations need to be installed at about 425~m from the
interaction region.
These detectors can have an acceptance in $\xi$ down to
1--2$\times 10^{-3}$, and a parametrization of the acceptance was
included in the event estimates in this paper.

In order to efficiently record and measure the diffractively
scattered protons in Roman Pot detectors or microstations, they
have to be sufficiently separated from the beam particles. The
detectors, which are  located at 420~m and 430~m from the
interaction point, could then be used to define the proton momenta
by measuring, with respect to the beam axis, the difference in
horizontal displacement at the two locations as a function of the
average proton deflection.

We observe that a variation of $\Delta \xi = 5 \times 10^{-4}$
produces a $80\:\mu{\rm m}$ difference in the horizontal
displacement of a diffractively scattered proton. With
state-of-the-art silicon microstrip detectors this difference can
be measured with a precision of the order of 5$\mu$m. The expected
momentum spread of the beam protons is $\Delta \xi/\xi = 10^{-4}$.
For a symmetric event configuration ($\Delta  =|\xi_1-\xi_2| \le
0.04$), we then expect in the most optimistic case a mass
resolution of the order of $\Delta M_{\rm missing}/M_{\rm
missing}$ better than 1\%~\cite{orava}. This leads to the value
$\Delta M_{\rm missing} = 1$~GeV, which is used in Table~1.

The acceptance of diffractively scattered protons is limited by
the minimum measurable deflection and, on the other hand, by the
aperture of the last dipole magnet (B11) of the LHC lattice, i.e.,
30mm, and leads to a range of observable missing masses of
$20\;{\rm GeV} < M_{\rm missing} <  160\:{\rm GeV}$. A similar
reason restricts the observable rapidity interval of the central
produced (Higgs) system. Since we do not measure $\xi$ smaller
than 1--2$\times10^{-3}$, we are unable to select events with
Higgs rapidity $|y_H|>$1.2--1.6. As a result, the efficiency of
proton tagging is 60\% for $M_H=120$~GeV. In calculating
acceptances, the detector edge effects (guard ring, shielding,
possible insulation layer, etc.) play a significant role and can
be further minimised. Effects due to detector alignment accuracy
have still to be assessed carefully.

For selecting the central diffractive events, a sufficient
suppression of the huge non-diffractive event rates has to be
provided. An event selection strategy could consist of the
following three steps:
\begin{itemize}
\item[(i)] At the first trigger level, select events with a pair of
jets in the central detector ($|\eta|< 2.5$) each with
$E_T>40$~GeV and with the difference between their azimuthal
angles $\phi_1 - \phi_2=180^\circ$ (within a given cell size of
$\Delta \eta \times \Delta \phi$). Furthermore, one can make use
of the absence of significant additional central activity in these
events.
\item[(ii)] At a later stage of data collection, or in off-line analysis, use the
forward-backward proton measurement to calculate the missing mass,
and
\item[(iii)] analyse the events of interest by using the
central detector data.
\end{itemize}

With condition~(i), a relative suppression of background to
central diffractive events of the order of $10^4$ is obtained.
Therefore, a well recordable first level trigger rate may be
achieved~\cite{orava}.

Pile-up events will also be important for the Roman Pot detectors.
The PYTHIA~\cite{PYTHIA} Monte Carlo program was used to estimate
the probability to have an additional proton accepted on one side
of the interaction region from single soft diffraction for the
different luminosities, and amounts to 8\% (medium luminosity),
40\% (high luminosity) and 200\% (SLHC\footnote{The Super LHC
luminosity is taken to be $10^{35}\:{\rm cm}^{-2}{\rm s}^{-1}$.}).
Hence at the SLHC special care is needed to control the
combinatorics generated by background from pile-up events. Since
by then the mass of the Higgs to some accuracy will be known, an
appropriate mass window can be chosen to select genuine scattered
protons that belong to the diffractive Higgs event.

The next issue is the efficiency $\varepsilon_b$ of tagging a $b$
jet. The value is correlated with the probability $P(g/b)$ to
misidentify a gluon as a $b$ jet. As we have seen in (i) of
Section~2.4.1, we require $P(g/b) = 0.01$ to reduce the $gg$
background to an acceptable level. For this value of $P(g/b)$, the
present estimate of the efficiency of $b$ and $\bar{b}$ tagging is
$(\varepsilon_b)^2=0.3$, but it is not inconceivable that this
could  be improved to a larger value, perhaps as large as
$(\varepsilon_b)^2=0.6$. If it turns out that this is impossible
for $P(g/b)=0.01$, then it is better to accept a worse
misidentification probability $P(g/b)$ in order to obtain a higher
value of $(\varepsilon_b)^2$. This will raise the background, but
will result only in a relatively small reduction in the
significance of the signal. For this reason we use
$(\varepsilon_b)^2 = 0.6$ in our estimates.

Therefore the event rate in (c) of Table~1 includes a factor 0.6
for the efficiency associated with proton tagging and 0.6 for $b$
and $\bar{b}$ tagging. Besides this the signal has been multiplied
by 0.5 for the jet polar angle cut and 0.67 for the $H\ra\bb$
branching fraction. Hence the original $(\sigma=3\:{\rm
fb})\times({\cal L}=30\:{\rm fb}^{-1})=90$ events is reduced to an
observable signal of 11 events, as shown in Table~1.

\section{Higgs production by Weak Boson Fusion (WBF)}

We have seen in Section~2 that the selection of events with large
rapidity gaps is an effective way of suppressing the QCD
background. Recall that rapidity gaps appear naturally in Weak
Boson Fusion (WBF) \cite{TDK,Bj}. One can thus exploit this
property to suppress the QCD background and to observe a light
Higgs boson produced by WBF via its main $H\ra\bb$ decay mode, in
addition to the rather rare decay modes, $\tau\tau$, $WW^\ast$,
etc. usually proposed for WBF, see Table~1 and Refs.
\cite{Z,WBF,WBFH}. Another special feature of Higgs production by
WBF is the high momentum transfer, $p_t\sim O(M_W)$. The hard
subprocess may be written as
\begin{equation}
qq\ra {\rm jet}\:W\:W\:{\rm jet}\ra{\rm jet} + H + {\rm
jet}\label{eq:qq}
\end{equation}
where again a + sign denotes a rapidity gap. The process is
sketched in Fig.~5.

\begin{figure}[!h]
\begin{center}
\epsfig{figure=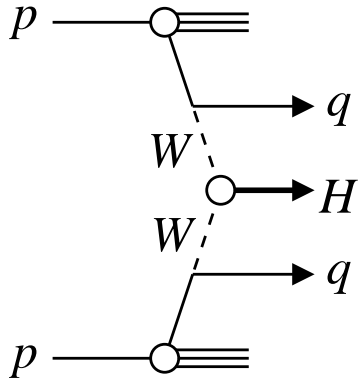,height=2.2in}
 \caption{Higgs production by weak boson fusion at the LHC, which is accompanied by forward and backward going quark jets.}
 \label{Fig.5}
\end{center}
\end{figure}

\subsection{The WBF $H\ra\bb$ signature}

Two alternative signatures of the WBF $H\ra\bb$ events exist.
First, we may select events with large $p_t$ (quark) jets in the
forward and backward directions, separated from the $H\ra\bb$
decay by rapidity gaps. Alternatively we may select events with a
high $q_t$ Higgs (or $\bb$ system), with rapidity gaps on either
side.

In practice the $H\ra\bb$ decay should be observed in the central
detector. We choose the rapidity $|y_{\rm Higgs}|<2.1$, and impose
the polar angular cut $60^\circ<\theta<120^\circ$ on the $b$ and
$\bar{b}$ jets (as discussed in Section~2.4). The accompanying
quark jets are observed in the forward and backward calorimeters.
For these we take\footnote{If we were to enlarge the coverage up
to $\eta_{\rm jet} = 7$ then we would increase the cross section
by only 3\%, leaving the signal-to-background ratio essentially
unaltered.} $3<|\eta_{\rm jet}|<5$. The results presented in the
upper entry of (f) in Table~1 correspond to the selection of
events with $p_t>30$~GeV (quark) jets in the forward and backward
directions, which are separated from $H\ra\bb$ by rapidity gaps.
For the alternative signature, shown by the lower entry of (f) in
Table~1, we impose the cut $q_t>30$~GeV on the $H\ra\bb$ system.
The detailed kinematics are described in Ref.~\cite{KMRhiggs}.

The leading order QCD background for process~(\ref{eq:qq}), with
$H\ra\bb$, is rather large, even after imposing the presence of
rapidity gaps and the high $p_t$ (or $q_t$) cuts. There is,
therefore, no need for a detailed discussion of the higher-order
background contributions in this case. Although $S/B$ is
relatively small, the cross section is considerably larger than
for the exclusive process and so the significance of the signal
looks good. In fact in some respects, this signal is similar to
the search for the decay $H\ra\gamma\gamma$ in an inclusive
process; compare entries~(a) and (f) of Table~1.

The main problem is that it is hard to identify rapidity gaps in
inclusive reactions at high luminosity, where pile-up becomes
significant. The possibility to identify the vertex of Higgs
production, and to separate off the particles associated with
other interactions, is discussed in Section~3.2.

The survival probability $S^2$ of the rapidity gaps to soft
rescattering is an ingredient in the calculation of both the
signal and the background of these $j+(\bb) +j$ events. It is
informative to discuss the values of $S^2$ that are obtained.
Recall that the survival of the gaps, at the hadronic level, are
computed using a two-channel eikonal model for soft rescattering
\cite{KMRsoft,KKMR}. For WBF the probability $S^2$ depends more
sensitively on the model than the previous calculation of $S^2$
for $pp\ra p + (\bb)+p$ of Section~2. The incoming partons which
participate in the WBF subprocess $qq\ra q + (\bb) + q$ are rather
hard. They have large $x$ and a larger scale. Thus in a
multi-channel eikonal approach, it is probable that such partons
belong to the component with the lower absorptive cross section.
Hence the survival probability $S^2$ is larger. Clearly the
results will depend on how the partons (of the global
analysis\footnote{We use MRST partons \cite{MRST}.}) are
distributed between the different diffractive eigenstates, that is
different channels of the eikonal.

Here we have used model~II of Ref.~\cite{KKMR}, which looks the
most realistic and which provides a good description of the CDF
data for the diffractive production of a pair of high $E_T$ jets.
For the first WBF signature (upper entry of (f) in Table~1), the
corresponding values of $S^2$ are 0.30 and 0.25 for the signal and
background respectively; whereas for the Higgs $q_t>30$~GeV
signature (lower entry) the values of $S^2$ are 0.26 and 0.21 for
the signal and background respectively.

In model~I of Ref.~\cite{KKMR}, which may be regarded as an
extreme case, it is assumed that all the valence quarks are
concentrated in the eikonal component with the smaller absorption
cross section and the gluons in that with the larger cross
section\footnote{This was the model used to generate the results
in the lower half of Table~1 of Ref.~\cite{KMRhiggs}.}. In this
case the QCD $\bb$ background, which originates mainly from
gluons, has gaps with a smaller survival probability ($S^2=0.09$
for both signatures (f) of Table~1) and the significance of WBF
Higgs signal would be increased to about $8\sigma$.

\subsection{Experimental issues associated with Higgs production by WBF with rapidity gaps}

Both the ATLAS and CMS experiments at the LHC prepare to measure
jets and energy flows as far out as $\eta = 5$,  matching the cuts
in Section~3.1. Hence the forward jets can be tagged and measured.
For the trigger typically four jet final states need to be
selected, supplemented with topological requirements, if one wants
to go down in jet $E_T$ threshold as much as 30~GeV.

When the LHC collider operates at medium and high luminosity, the
recorded events will be plagued by overlap interactions in the
same bunch crossing. At medium luminosity (i.e., $10^{33}\:{\rm
cm}^{-2}{\rm s}^{-1}$) on average 2.3 inelastic events are
expected to be produced in each bunch crossing. Hence the rapidity
gaps will often be destroyed by these additional pile-up events.
In particular, for the high luminosity operation ($10^{34}\:{\rm
cm}^{-2}{\rm s}^{-1}$), on average 22 additional events are
overlaid on top of the signal event, which will essentially always
destroy the gap.

It is, however, possible to use the detector information to try to
reconstruct the gap in the hard scattering events. The vertices of
the individual collisions will be (non-uniformly) distributed
along the beam axis in the interaction region over a distance of
10--20~cm. The precise tracking subdetectors of the experiments
will, however, allow the reconstruction of vertex positions with a
precision of a few tens of microns, and even soft tracks can be
associated to their corresponding vertex with a precision of a
fraction of a millimeter. Thus one can imagine an event selection
that checks for rapidity gaps based on the charged particles
associated with the proper vertex. Furthermore, the transverse
energy of particles from the soft overlap events is generally low
and, for example, considering only particles with an  $E_T$ value
larger than of order 1--2~GeV will often reveal the underlying
rapidity gap of the hard scattering event.

All rapidity gaps in the events should be detectable by vertex
and/or soft energy cuts particularly for the data taken at medium
luminosity, and probably also for high luminosity data samples.
However, it is unlikely that these techniques can be used for the
SLHC type of luminosity of $10^{35}\:{\rm cm}^{-2}{\rm s}^{-1}$.

\section{Discussion}

\subsection{Measurement of Higgs couplings}

As emphasized in the Introduction there is no single obvious best
discovery channel for a light Higgs boson at the LHC. Rather the
search should employ all possibilities. The comparability of
channels has some advantages in the measurement of Higgs
couplings. The Double-diffractive Higgs production processes
(entries (c) and (d) of Table~1) are mediated by the subprocess
$gg\ra t\bar{t}\ra H$, and are thus proportional to the
$H$--$t\bar{t}$ coupling (squared). The same is true for
production processes (a) and (b). For process (a) the
$H\ra\gamma\gamma$ decay is mainly controlled by the $H$--$WW$
coupling, while processes (b) and (c) are proportional to the
$H$--$\bb$ coupling (squared). From Table~1 we note the relatively
large significance of the WBF processes, (e) and (f), in which the
Higgs signal is separated from forward and backward high $p_t$
jets by rapidity gaps. These WBF processes are, of course, driven
by the $H$--$WW$ coupling. If the $H\ra\tau\tau$ decay process is
observed, then we can study the Higgs-lepton coupling. The
estimates presented in Table~1 for this $qq\ra qHq\ra j\tau\tau j$
process are taken from Ref.~\cite{Z}. The values agree reasonably
well with the results\footnote{The $\tau\tau$ results in Table~1
of \cite{KMRhiggs} included a 10\% efficiency to identify the two
$\tau$ leptons.} shown in Table~1 of Ref.~\cite{KMRhiggs}. Thus by
measuring combinations of the above processes it is possible to
determine the Higgs coupling to $b$ and $t$ quarks, $W$ bosons and
the $\tau$ lepton.

\subsection{Conclusion}

This paper has concentrated on the production of a light Higgs
boson accompanied by rapidity gaps. In Section~2 we discussed
exclusive double-diffractive (CP-even) Higgs production
\begin{equation}
pp\ra p+H(\bb)+p, \label{eq:pp}
\end{equation}
and in Section~3, Weak Boson Fusion via the subprocess
\begin{equation}
qq\ra q+H+q \ra {\rm jet}+\bb + {\rm jet}.\label{eq:qqconclusion}
\end{equation}
In both cases we estimated the cross section for the $H\ra \bb$
signal and for the QCD $\bb$ background, at the LHC. The results
obtained are summarised in entries (c) and (f) of Table~1.

Provided that appropriate proton taggers are installed (see
Section~2.6), process (\ref{eq:pp}) has the special advantage that
the Higgs can be identified by a sharp peak in the protons'
`missing mass' spectrum and, simultaneously, as a peak in the
$\bb$ mass spectrum. The required equality $M_{\rm missing} =
M_{\bb}$, allowing for resolution, is of great value, not only to
establish the signal, but also to suppress the $\bb+ng$
background. In addition, the existence of a $J_z=0$ selection rule
automatically greatly suppresses the leading order (LO) QCD $\bb$
background. We estimated the LO, NLO and NNLO contributions to the
$\bb+ng$ background. It turned out that, for production from a
colour-singlet two-gluon system, the special colour and helicity
structure of the subprocess is such that even the higher order
(NLO, ...) contributions to the background are suppressed. In
summary, we find\footnote{Here, and in Table 1, we
(conservatively) assume that the higher-order virtual contribution
to the background has approximately the same relative size (that
is the same $K$ factor) as the signal.}
\begin{equation}
\frac{\rm Signal}{\rm Background} \simeq 3\label{eq:StoB4}
\end{equation}
for the exclusive double-diffractive production of a light Higgs
boson, decaying via $\bb$, at the LHC. The favourable
signal-to-background ratio is offset by a low event rate, caused
by the necessity to preserve the rapidity gaps so as to ensure an
exclusive signal. Nevertheless, entry~(c) of Table~1 shows that
the signal has comparable significance\footnote{However, when the
NLO contributions to the $H\ra\gamma\gamma$ signal and background
are included, the significance of this signal is increased to
about $7\sigma$ for 30~fb$^{-1}$ LHC luminosity and Higgs mass
$M_H\simeq 120\:{\rm GeV}$~\cite{BDS}.} to the standard
$H\ra\gamma\gamma$ and $t\bar{t}H$ search modes (entries~(a) and
(b)).

The cross section for the production of a 120~GeV Higgs boson at
the LHC, via the exclusive $pp\ra p+H+p$ process, was calculated
to be 3~fb (see Section~2.1), but after including the $H\ra\bb$
branching fraction, and the acceptance and efficiency cuts, we
arrive at only 12\% of the signal. The breakdown of the depletion
of the signal is summarized at the end of Section~2.6. Thus for an
integrated luminosity of 30~fb$^{-1}$ (300~fb$^{-1}$) we would
register 11 (110) events. Noting the $B/S$ ratio of
(\ref{eq:StoB4}), we see that these signals have a significance of
about $3\sigma$ and $9\sigma$ respectively. We estimate a factor
of two uncertainty in the cross section for this exclusive Higgs
signal (see Section~2.2), but a much better reliability for the
signal-to-background prediction, (\ref{eq:StoB4}), since the main
theoretical uncertainties cancel in the ratio (see Section~2.3).

The Weak Boson Fusion signal, (\ref{eq:qqconclusion}), does not
need the installation of proton taggers, and has a favourable
significance; see Section~3 and Table~1. As discussed in
Section~3.2, the main problem is to identify these rapidity gap
events from the additional pile-up events, that is from overlap
interactions in the same bunch crossing. There is good reason to
believe that, at the medium luminosity of $10^{33}\:{\rm
cm}^{-2}{\rm s}^{-1}$ of the LHC, the problem can be overcome, and
that we can go a considerable way to achieving the numbers quoted
in entry~(f) of Table~1.

In conclusion, we have shown how the rapidity gap processes,
(\ref{eq:pp}) and (\ref{eq:qqconclusion}), may play a key role in
identifying and studying a light Higgs boson at the LHC.

\section*{Acknowledgements}

We thank Andrei Shuvaev for valuable help in the calculations of
the QCD $\bb$ background, and Marco Battaglia, Michael Kraemer,
Jerry Lamsa, James Stirling, Theodore Todorov and Peter Williams
for useful discussions. One of us (VAK) thanks the Leverhulme
Trust for a Fellowship. This work was partially supported by the
UK Particle Physics and Astronomy Research Council, by the Russian
Fund for Fundamental Research (grants 01-02-17095 and 00-15-96610)
and by the EU Framework TMR programme, contract FMRX-CT98-0194 (DG
12-MIHT).




\vfill

\newpage


\end{document}